\documentclass[a4paper,fleqn,usenatbib]{mnras}
\usepackage{newtxtext,newtxmath}
\usepackage[T1]{fontenc}
\usepackage{ae,aecompl}

\usepackage{graphicx}	
\usepackage{amsmath}	
\usepackage{amssymb}	

\usepackage[caption=false]{subfig}
\usepackage{epstopdf}
\usepackage{savesym}

\title[Evolution of Illustris Starbursts]{Evolution of Starburst Galaxies in the Illustris Simulation}

\author[C. L. Wilkinson et al.]{
C. L. Wilkinson,$^{1}$\thanks{E-mail: c.wilkinson@2014.hull.ac.uk}
K. A. Pimbblet,$^{1}$
J. P. Stott,$^{2}$
C. G. Few,$^{1}$
B. K. Gibson$^{1}$
\\
$^{1}$E.A. Milne Centre for Astrophysics, University of Hull, Cottingham Road, Kingston Upon Hull, HU6 7RX, UK\\
$^{2}$Department of Physics, Lancaster University, Lancaster LA1 4YB, UK
}

\date{Accepted XXX. Received YYY; in original form ZZZ}

\pubyear{2018}

\begin{document}
\label{firstpage}
\pagerange{\pageref{firstpage}--\pageref{lastpage}}
\maketitle

\begin{abstract}
There is a consensus in the literature that starburst galaxies are triggered by interaction events. However, it remains an open question as to what extent both merging and non-merging interactions have in triggering starbursts? In this study, we make use of the Illustris simulation to test how different triggering mechanisms can effect starburst events. We examine star formation rate, colour and environment of starburst galaxies to determine if this could be why we witness a bimodality in post-starburst populations within observational studies. Further, we briefly test the extent of quenching due to AGN feedback. From Illustris, we select 196 starburst galaxies at $z$~=~0.15 and split them into post-merger and pre-merger/harassment driven starburst samples. We find that 55\% of this sample not undergone a merger in the past 2~Gyr. Both of our samples are located in low-density environments within the filament regions of the cosmic web, however we find that pre-merger/harassment driven starburst are in higher density environments than post-merger driven starbursts. We also find that pre-merger/harassment starbursts are redder than post-merger starbursts, this could be driven by environmental effects. Both however, produce nuclear starbursts of comparable strengths.
\end{abstract}

\begin{keywords}
galaxies: evolution -- interactions -- starburst -- star formation
\end{keywords}



\section{Introduction}
The star formation main sequence (SFMS) is a tight correlation (scatter $\approx$ 0.2~dex as reported by \citealt{Speagle2014}) between mass and star formation rate (SFR). It holds true at both low (\citealt{Brinchmann2004,Salim2007}) and high redshifts (\citealt{Daddi2007}) over multiple wavelengths (\citealt{Elbaz2011,Rodighiero2014}). Star-formation occurs in two modes: quiescently and star bursting (\citealt{Pillepich2017}), as revealed by the Kennicutt-Schmidt relation (\citealt{Kennicutt1998}). Most galaxies situated in the SFMS can be considered star forming at a steady rate, whilst those significantly above the main sequence are considered to be in a starburst phase. Galaxies undergoing starburst spend a very short amount of time in this phase ($\sim$10$^8$~years) and because of this, they are rare, making up between only 5\% and 10\% of the global galaxy population (\citealt{Rodighiero2011}).

Throughout the literature there is a consensus about the possible triggers of starburst/post-starburst galaxies, namely mergers (\citealt{Barnes1991,Sparre2016}) and galaxy-galaxy interactions (\citealt{Zabludoff1996}). It is suggested that these mechanisms are responsible for transforming star forming spirals into quiescent ellipticals. Evidence of this can be seen in the morphologies of post-starburst galaxies. \cite{Zabludoff1996} finds that in a sample of 21 post-starburst galaxies from the Las Campanas Redshift Survey that 5 galaxies display tidal features. This is a consequence of galaxy interactions. Other studies such as \cite{Tran2004} and \cite{Quintero2004} find that the morphology of post-starburst galaxies are generally bulge dominated with underlying disc components, similar to the S0 morphology, reinforcing the evolutionary link between spiral and elliptical galaxies.

It is widely believed that major mergers contribute heavily to the production of elliptical galaxies (\citealt{Cox2006}). Using 112 $N$-body merger simulations of varying mass ratios, \cite{Naab2003} find that mergers with mass ratios of 1:1 - 1:4 mostly result in elliptical-like remnants. These remnants can be discy and resemble an S0 morphology, similar to the post-starburst morphologies found by \cite{Yang2004}. Work by \cite{Sparre2017a}, using the Illustris simulation (\citealt{Genel2014,Vogelsberger2014,Vogelsberger2014b}), find that merger remnants are able to regrow their disc and do not necessarily have to be quenched ellipticals.

However, spiral morphologies are found in 44--54\% of post-starburst galaxies (\citealt{Wilkinson2017a}), meaning major mergers are not the only process triggering the starburst phase. \cite{Cales2013} suggests that non-merging galaxy interactions are more likely to maintain a spiral morphology. This could suggest why there is a significant fraction of post-starburst spiral galaxies. However, this is not the only alternative to mergers to trigger starbursts, other studies such as \cite{Dekel2009a}, \cite{Ceverino2010}, \cite{Cacciato2012} and \cite{Porter2014} suggest that instabilities in the disc could also result in starbursts.

Further, the environment in which starburst galaxies are found supports the merger/galaxy-galaxy interaction connection. Mergers, particularly gas-rich major mergers, are found to be more prominent in low-density environments (\citealt{Bekki2001c, Lin2010, Sanchez-Blazquez2009a}). These results back up the findings by \cite{Hashimoto1998}, who find that star formation is on average higher in field environments than in cluster environments. This in turn reveals that there is a higher fraction of starbursts in the field compared with clusters. \cite{Zabludoff1996} finds that 75\% of post-starburst galaxies are located in the field, a similar result is found in \cite{Wilkinson2017a}. By combining these findings, it suggests that for the majority of starburst galaxies major mergers are the main trigger (\citealt{Wild2009c, Snyder2011a}) but not the only trigger (\citealt{Sparre2017a}) for driving galaxy transformation, with galaxy-galaxy interactions being the next major trigger.

Whilst starbursts are twice as common in the field than in cluster environments (\citealt{Poggianti1999}), they are present in some cluster environments (\citealt{Balogh1999}). \cite{Poggianti2009} find that post-starburst galaxies are predominantly in cluster environments at higher redshifts (0.4 < $z$ < 0.8). At lower redshifts (0.02 < $z$ < 0.06), \cite{Mahajan2013h} find that post-starburst galaxies prefer a weak-group environment containing 4 to 10 group members and that 86\% of X-ray bright clusters contain sub-structure on the weak group scale. Of these weak groups, 91\% contain post-starburst galaxies. This suggests pre-processing is occurring, in which the starburst is triggered in a weak group environment which then infalls into a denser cluster environment. This results in ram pressure stripping that quenches star formation.

In both cases, mergers and interactions have the potential of triggering a starburst (\citealt{Zabludoff1996,Bekki2001c,Bekki2005a,Hopkins2006a,Hopkins2008}) in which tidal torques funnel gas into the galactic centre (\citealt{Barnes1991}; \citealt{Barnes1996}). The increased build up of gas in the galactic centre then begins to fuel a rapid burst of star formation known as a nuclear starburst. \cite{Sparre2016} suggests that head-on mergers are likely to produce a strong nuclear burst were the strength of the burst is directly proportional to the speed of the collision. Due to the regulatory processes within galaxies, this elevated rate of star formation is not sustained for a prolonged period of time and is quenched.

The literature suggests many potential mechanisms that could quench star formation after a starburst such as AGN feedback, stellar feedback, ram pressure stripping, and gas depletion. AGN feedback is a form of rapid quenching and has been discussed extensively within the literature (\citealt{Springel2005b, Goto2006d, Feruglio2010, Cicone2014}) and is typically attributed to gas-rich major mergers (\citealt{DiMatteo2005, Hopkins2006a}). When the merger occurs, gas is funnelled into the centre of the galaxy and activates the AGN, causing `quasar mode' feedback which ejects remaining gas away from the star forming central region via strong galactic winds. After the quasar phase, `radio mode' feedback takes over in which the AGN heats up surrounding gas preventing it from forming stars (\citealt{Croton2006}). Whilst AGN feedback is a powerful tool in suppressing star formation, gas falling back from the initial blow-out is capable of reigniting star formation (\citealt{Faucher-Giguere2018}).

Star-formation triggered by either minor mergers or galaxy-galaxy interactions are found to quench on an intermediate time-scale, 1.0 $\lesssim$ $\tau$/Gyr $\lesssim$ 2.0 (\citealt{Smethurst2015}). Mechanisms that could quench on an intermediate time-scale include ram pressure stripping, gas depletion or harassment. The morphology of such a remnant would resemble an S0 morphology, similar to the post-starburst galaxies found by \cite{Tran2004}, \cite{Quintero2004} and \cite{Yang2004}.

There has been many studies focusing on the link between starburst and mergers. However, there is little investigation in the literature that examines the links between non-merging events and starburst and how they compare to post-merger starbursts. In this study we use hydrodynamical simulations from Illustris (\citealt{Vogelsberger2014,Nelson2015}) to track the evolution of starburst galaxies. We explicitly aim to determine their main trigger and make a comparison between post-merger and pre-merger/harassment driven starbursts.

In section 2 we give a brief description of the Illustris simulation and how we derive our sample. In section 3 we discuss our findings on the triggering mechanisms of starburst galaxies, their properties and what quenches their star formation. In section 4 we will discuss our main findings and in section 5 we will make our conclusions.

\section{Sample Selection}

\subsection{Illustris}
In order to address the questions above, we use the Illustris simulation to track and compare starbursts driven by differing triggers. Illustris is a hydrodynamical simulation that tracks cosmological evolution from $z$ = 127 to $z$ = 0 in a box of comoving size 106~Mpc$^3$ (\citealt{Genel2014, Vogelsberger2014, Vogelsberger2014b}). The following cosmological parameters are adopted: $\Omega_m$ = 0.2726, $\Omega_{\Lambda}$ = 0.7274, $\Omega_b$ = 0.0456, $\sigma_8$ = 0.809, $n_s$ = 0.963 and $H_0$ = 100~$h$~km~s$^{-1}$~Mpc$^{-1}$ where $h$ = 0.704. The initial conditions are generated at $z$ = 127 and achieves a dark matter resolution of 6.26~$\times$~10$^6$~M$_{\odot}$ and baryonic matter mass resolution of 1.26~$\times$~10$^6$~M$_{\odot}$. The smallest radii of a cell achieved is 48~pc (\citealt{Vogelsberger2014}).

Illustris uses the moving-mesh code \textsc{arepo} (\citealt{Springel2010a}) that provides a hydrodynamical treatment of gas and works alongside gravitational forces (calculated using a Tree-PM scheme; \citealt{Xu1995}) to create realistic galaxy formation. Phenomenological models are also included to allow for processes that regulate stellar mass growth within galaxies such as AGN feedback, stellar mass loss and SMBH growth. The simulation is capable of resolving gravitational dynamics down to ranges of 710~pc at $z$ = 0 whilst following large-scale evolution (\citealt{Vogelsberger2014}).

Illustris uses the star formation and feedback model from \cite{Springel2003}. The model describes the multiphase nature of star formation, accounting for both self-regulating, `quiescent' star formation and `explosive' star formation. \cite{Springel2003} use a sub-resolution model that uses spatially averaged properties to describe the ISM; this includes the growth of cold gas clouds, radiative cooling and supernova feedback in the form of galactic winds, radiative heating and outflows. During a starburst the gas density is much higher than the star formation threshold and this allows for efficient star formation. However, in the self-regulating model, galactic winds can reduce the efficiency of star formation producing results that are consistent with observations.

We make use of the publicly available merger trees (\citealt{Nelson2015}) created using the \textsc{sublink} code (\citealt{Rodriguez-Gomez2015}). These trees allow us to track our selected galaxies through subsequent and previous time steps and give details about their merger histories. This will allow us to compare the driving forces behind starburst galaxies.

\subsection{Starburst Selection}
We begin by selecting starburst galaxies at a lookback time of 1.912~Gyr ($z$ = 0.15) with the aim of tracking them forward through time to $z$ = 0. By starting at this point in the simulation we are able to track the galaxies through 12 snapshots to $z$ = 0 which covers the 2~Gyr duration of the post-starburst phase (\citealt{Kaviraj2007a}). Selecting galaxies in one snapshot simplifies the analysis as the time between snapshots is not uniform. Working in a low-$z$ regime allows for the minimum time difference between snapshots, allowing us to monitor the evolution of low-$z$ galaxies more closely and provides ease of comparison to SDSS studies like \cite{Wilkinson2017a}.

\begin{figure}
	\centering
	\includegraphics[width=0.5\textwidth]{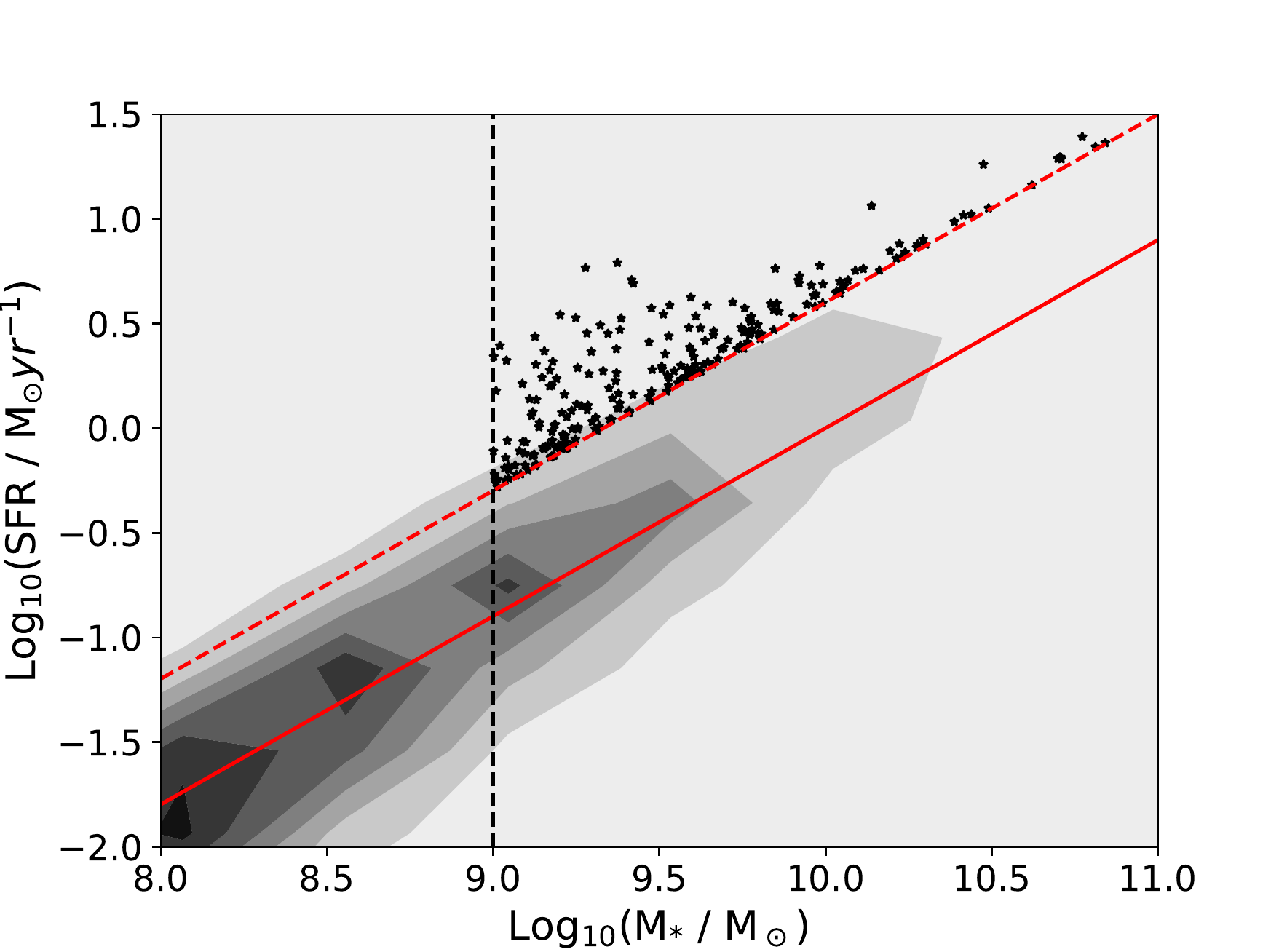}
	\caption{This plot shows the star formation main sequence at a lookback time of 2~Gyr. The red solid line denotes the star formation main sequence fitted using linear regression. We select starburst galaxies above the red dashed line 0.6~dex above the main sequence. We select galaxies with masses above 10$^9$M$_{\odot}$. Using this criteria we select 196 starburst galaxies.}
	\label{Starbursts}
\end{figure}

To select starburst galaxies we first plot the star formation main sequence, as shown in Fig.~\ref{Starbursts}.  We exclude passive galaxies at $z$ = 0.15 from our fit by fitting our trend line to main-sequence galaxies with specific star formation rates above 2.5~$\times$~10$^{10}$~yr$^{-1}$. Further, we apply a minimum mass of 10$^{9}$~M$_{\odot}$ for ease of comparison to observational studies. We identify galaxies 0.6~dex above the main sequence line as starburst galaxies, as defined by \cite{Zhang2016}. Using this method, we select 196 starburst galaxies (0.82\% of Illustris galaxies in this snapshot within the mass range given above).

We add the caveat that within the Illustris simulation, starbursts are under-produced (\citealt{Sparre2015}), this is due to the spatial resolution of the simulation (\citealt{Sparre2016}). This problem reduces the star formation that is measured and therefore hides the bursty nature of starbursts. This could cause some starbursts to be hiding within the main sequence. The star formation main sequence becomes less defined at higher masses and as a result the average star formation rate drops as masses surpass 10$^{10.5}$~M$_{\odot}$ (\citealt{Sparre2015}). This results in fewer starburst galaxies being identified at higher masses.

\section{Results}
\subsection{Triggering Mechanisms of Starbursts}

\begin{table}
	\centering
	\begin{tabular}{c|c|c|c|c|c}
		\hline
		&& \multicolumn{4}{c}{log$_{10}$(M/M$_{\odot}$)}\\
		\hline
		Sub-Sample & Total & 9.0-9.5 & 9.5-10.0 & 10.0-10.5 & 10.5-11.0\\
		\hline
		Major & 35 & 20 & 12 & 2 & 1\\
		Intermediate & 24 & 5 & 12 & 5 & 2\\
		Minor & 30 & 9 & 15 & 4 & 2\\
		Post-Merger & 89 & 31 & 38 & 11 & 5\\
			\begin{tabular}{@{}c@{}}Pre-Merger/ \\ Harassment\end{tabular}
		& 107 & 58 & 31 & 15 & 3\\
		\hline
	\end{tabular}
	\caption{Here, we explore the merger histories of our starburst sample by determining how many have had a merger in the past 2~Gyr and the mass ratios of such mergers. The mass ratios are as follows: 1:1-1:4 (major), 1:4-1:10 (intermediate) and 1:10-1:100 (minor). We also separate our findings by mass. We find that just over half of the starburst galaxies in this sample (55\%) have not had a merger in the previous 2~Gyr.}
	\label{merger histories}
\end{table}

\cite{Zabludoff1996} suggested that the main mechanisms responsible for triggering the starburst phase are galaxy-galaxy interactions and mergers. Studies such as \cite{Wild2009b} and \cite{Snyder2011a} suggest gas rich major mergers are responsible for triggering starburst galaxies. Other studies have reported non-merging galaxy-galaxy interactions are more likely to maintain spiral structure post starburst (\citealt{Cales2013, Wilkinson2017a}). There are an extensive number of studies researching the link between mergers and starbursts but to what extent do non-merging interactions play on triggering starbursts?  In this section we explore the potential triggers of starbursts by making use of the \textsc{sublink} merger trees in Illustris.

We examine the merger histories of our starburst galaxies and split our primary sample into two sub-samples based on whether they have undergone a merger in the previous 2~Gyr. Those starbursts that have had a merger in the past 2~Gyr we call post-merger starbursts and those that have not we call pre-merger/harassment starbursts, we note that this sample may or may not have a harassing neighbour that could or could not lead to a merger. However, at the snapshot in which these galaxies are selected, there has been no coalescence in the previous 2~Gyr and it is in this way that our two samples differ. We explore the post-merger scenario further by examining the types of merger that have occurred by splitting our post-merger driven starburst sample by mass ratio; 1:1-1:4 (major mergers), 1:4-1:10 (intermediate mergers) and 1:10-1:100 (minor mergers) as shown in Table~\ref{merger histories}. Mass ratio is defined as the ratio of stellar mass of the merger, calculated at a time when the secondary progenitor reaches its maximum stellar mass (\cite{Rodriguez-Gomez2015,Rodriguez-Gomez2016}). We find that 35 (15\%) have had a major merger, as defined by \cite{Bournaud2005}. We use this definition as it has been shown to produce remnants with similar morphologies to the post-starburst galaxies found by \cite{Naab2003} and \cite{Tran2003a}.

We separate out intermediate mergers (those with ratios between 1:4 and 1:10) because \cite{Bournaud2005} finds that this type of merger can form remnants with S0 morphologies. We find that 24 galaxies (10\%) in our sample have had this type of merger in the previous 2~Gyr. When examining minor mergers, those with mass ratios less than 1:10, we find that 30 starbursts (34\%) have had minor mergers, i.e. those with ratios less than 1:10 in the past 2~Gyr. We also split these fractions by mass in Table~\ref{merger histories}, that shows minor mergers are more prevalent at higher mass regimes whilst major mergers are more likely to occur at lower mass regimes because the number density of galaxies drops as mass increases.

In total 89 (13\%) starburst have had at least one merger in the past 2~Gyr which suggests mergers do have a significant impact on triggering starbursts. However, mergers are not the only trigger as suggested by \cite{Sparre2017}; what is triggering the remaining 107 (62\%) starbursts? Perhaps, another likely trigger is harassment interactions which can cause rotational instabilities that lead to tidal torques capable of tunnelling gas and dust into the galactic centre \cite{Barnes1992}.

To determine how starbursts in the pre-merger/harassment sample are being triggered, we examine the locations and track the movements of galaxies within a 100~kpc radius surrounding the pre-merger/harassment starburst sample. We find that 52 pre- merger/harassment starbursts ($\sim$49\% of the pre-merger/harassment starburst sample) have a neighbour with a stellar mass at least 10\% that of the starburst galaxy. This could indicate that harassment events are triggering starbursts whether it be a pre-merger harassment or a pure harassment event with no consequent merger. By examining the merger trees for future mergers, we find that 33 starbursts (31\%) in our pre-merger/harassment sample have a future merger with a minimum mass ratio of 1:10 in subsequent snapshots. However, without the full raw data from Illustris we are unable to get a quantitative measure of tidal gravity. In 8 ($\sim$7\%) of the pre-merger/harassment starburst galaxies, there are no surrounding galaxies or  mass. In these galaxies, gas could be accreted from the intergalactic medium fuelling a starburst or instabilities in the galactic disc could be driving the rapid burst of star formation. Only a higher resolution simulation would allow us to determine what is causing burst of star formation. These results could suggest that interactions play a role in triggering starbursts.

\begin{figure}
	\centering
	\includegraphics[width=0.5\textwidth]{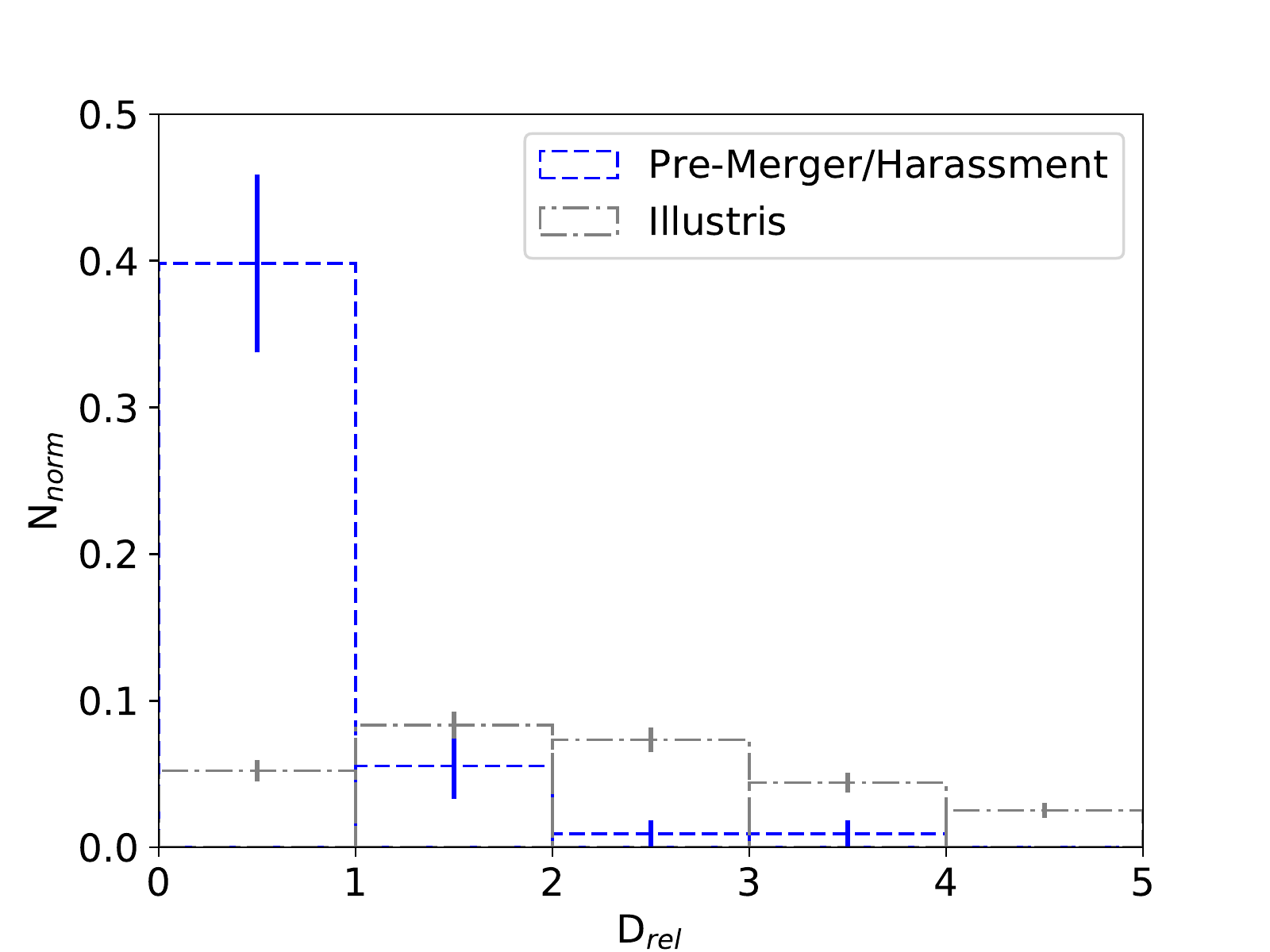}
	\caption{The distribution of relative distances of pre-merger/harassment starbursts and Illustris control galaxies to their closest neighbour, normalised to their relative total sample size. For $\approx$40\% of the pre-merger/harassment starbursts, they have a neighbour within a relative distance of 1. This means that for these galaxies, the distance between the two galactic centres is less than the sum of their radii.}   
	\label{rel_dist}        
\end{figure}

For starbursts in the pre-merger/harassment sample that have neighbouring galaxies within a 100~kpc radius and are greater than 10\% of the starburst mass, we investigate the harassment scenario by determining the relative distance to the closest neighbour. Fig.~\ref{rel_dist} shows the distribution of relative distances normalised to the total sample size. We define relative distance, \textit{D$_{rel}$}, below in Eq. \ref{eq_reldist}, where \textit{D} is the distance from the centre of the starburst galaxy to the centre of its closest neighbour, \textit{R$_1$} is the half mass radii of the starburst and \textit{R$_2$} is the half mass radii of the closest neighbour. This method allows us to determine how close galaxies get regardless of their size.

\begin{equation}
D_{rel}~=~D~/~(R_1~+~R_2)
\label{eq_reldist}
\end{equation}

In Fig.~\ref{rel_dist} we include a non-starburst control sample that is composed of 500 randomly selected galaxies from the Illustris simulation with masses above 10$^9$~M$_{\odot}$. We find that $\sim$40\% of the pre-merger/harassment starburst sample have a neighbour with \textit{D$_{rel}$} $<$~1 which quickly drops to $\ll$10\% for neighbours with relative distances greater than 1. Fig.~\ref{rel_dist} shows the control sample peaks at a higher \textit{D$_{rel}$} values and has a much lower fraction at \textit{D$_{rel}$} $<$~5 than our pre-merger/harassment starburst sample. This shows that starbursts are closer to their neighbours than non-starbursts. This is strong evidence for an interaction driven starburst. Due to the close proximity it is not unreasonable to predict that a merger will eventually follow, providing that the relative velocities are low enough that will allow the galaxies to coalesce.

\subsection{Environments}
As we find in our previous study (\citealt{Wilkinson2017a}), post-starburst galaxies predominantly reside in, but not restricted to, low-density environments. We find that those in high-density environments are redder and more elliptical in morphology than those in low-density environments. This suggests that environment could be enhancing the evolution in clusters and rich-groups. To discover more about the effects of environment on the post-starburst phase we look at the locations of our starburst galaxies in the Illustris simulation.

\begin{figure}
	\centering
	\includegraphics[width=0.52\textwidth]{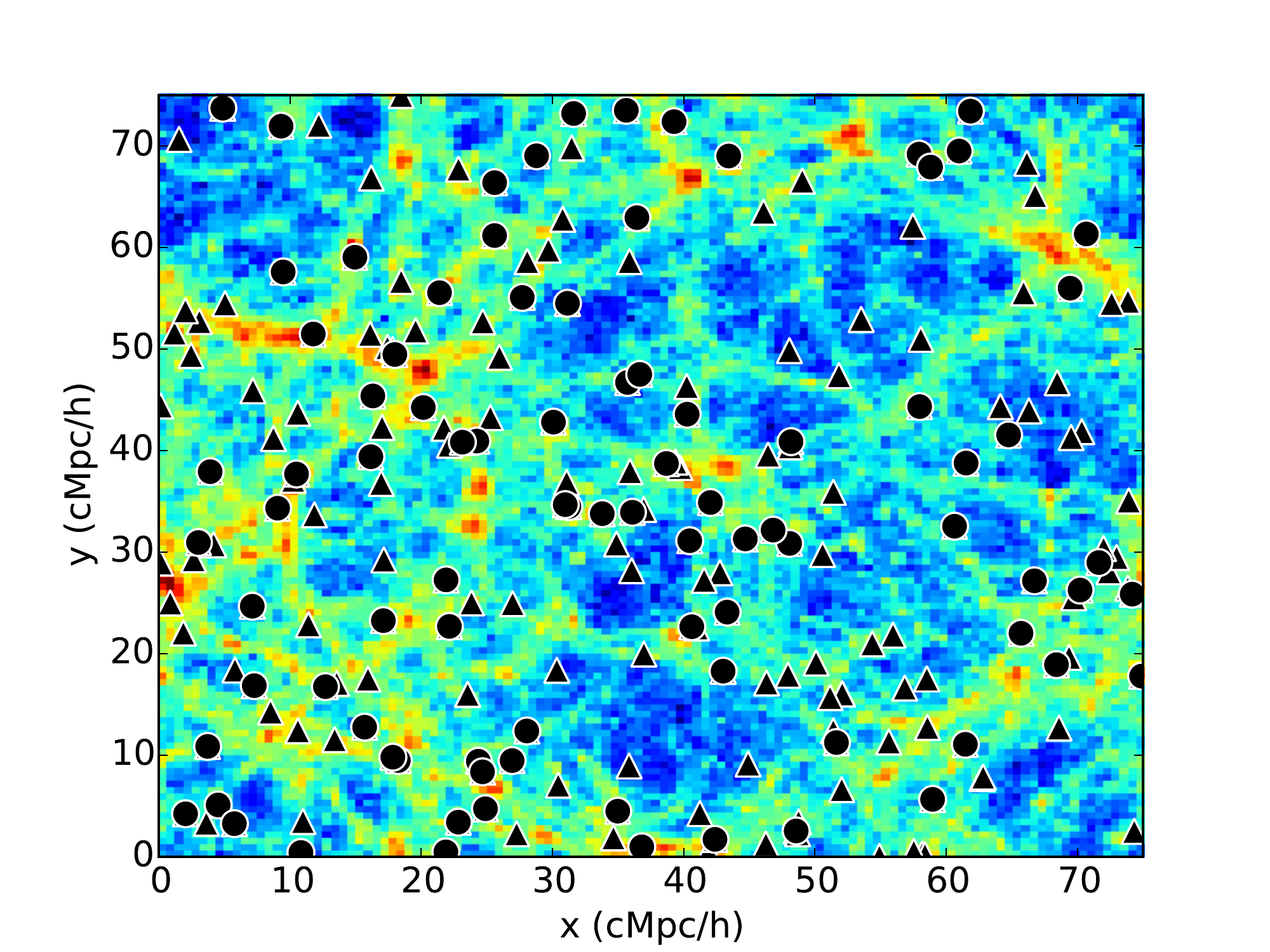}
	\caption{A 2D view of the Illustris simulation highlighting the locations of starburst galaxies with mergers in the previous 2 Gyr (circles) and pre-merger/harassment starbursts (triangles). The redder regions of the plot represent the densest areas within Illustris, whilst the blue regions are the least dense. We find that starbursts are predominantly in low-density regions around filaments within the cosmic web.}
	\label{global_env}
\end{figure}

Firstly, we explore the global environments of starburst galaxies in the Illustris simulation. Fig.~\ref{global_env} shows the locations of starburst galaxies against the number density of galaxies in Illustris. We highlight the locations of starbursts that have had some sort of merger in the last 2~Gyr. We can see that the majority of the starbursts are located in the lower density environments around the filament regions.

\begin{figure*}
	\centering
	\subfloat{
		\includegraphics[width=0.5\textwidth]{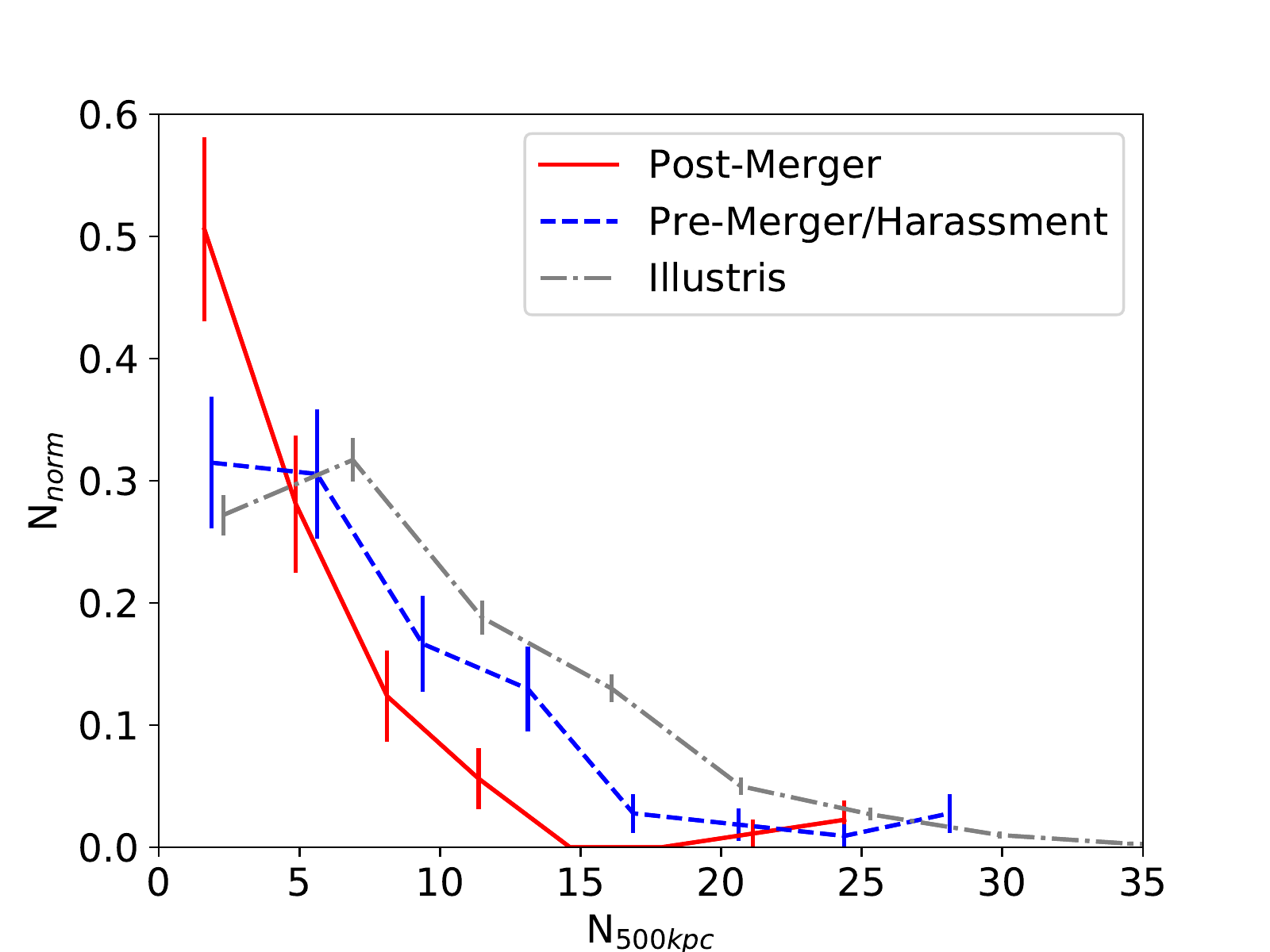}
	}
	\subfloat{
		\includegraphics[width=0.5\textwidth]{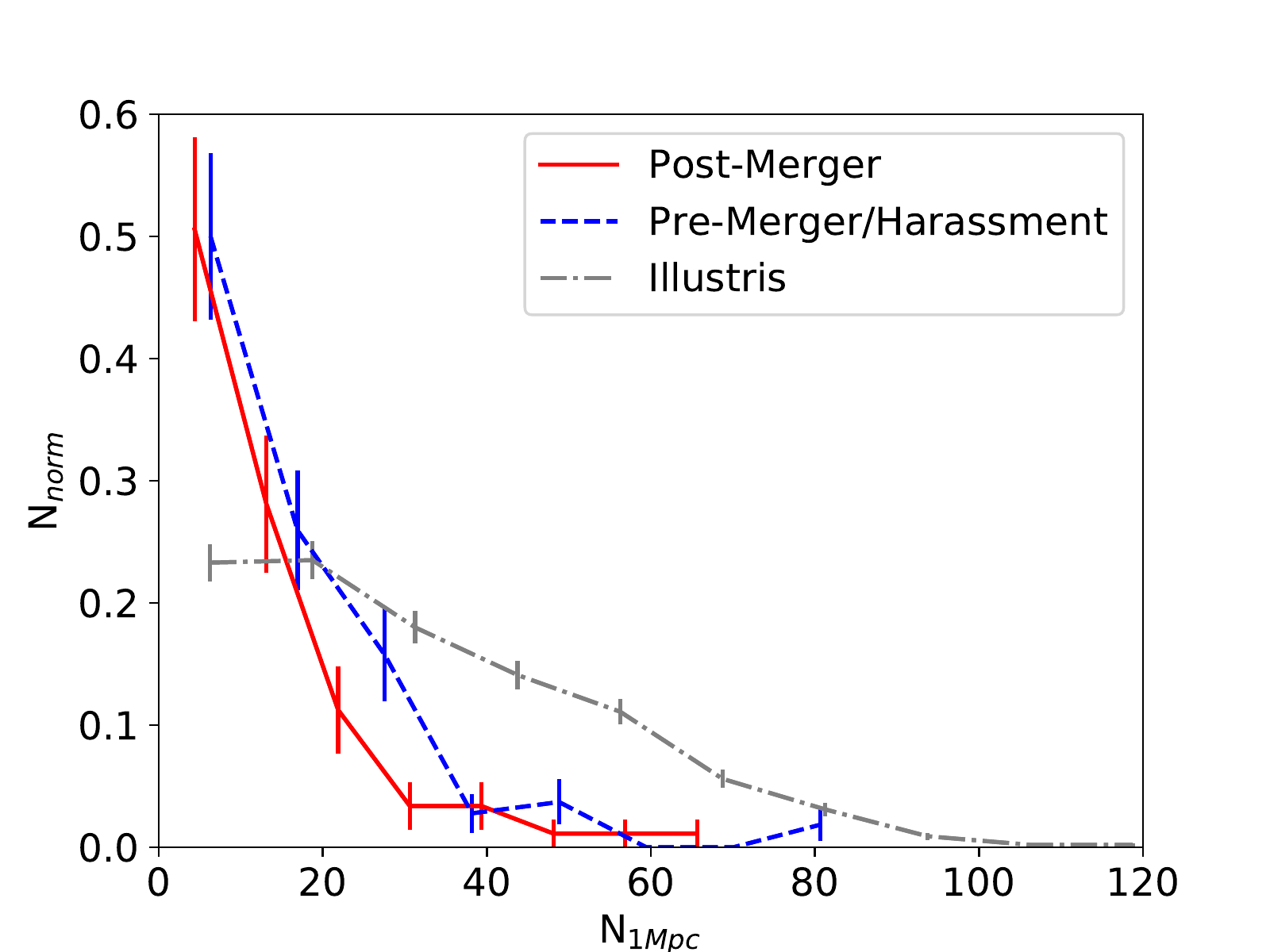}
	}
	\caption{These plots show the number density of galaxies with a minimum mass of 10$^5$M$_{\odot}$ (this is the minimum stellar mass of a resolved subhalo) surrounding our samples within a volume of radii 500~kpc and 1~Mpc. We see that the pre-merger/harassment starburst sample (blue) has an extended tail which would indicate their locations in higher density environments, whilst merger driven starbursts reside in much weaker environments. In both plots, control galaxies from Illustris are in denser environments than both starburst samples.}
	\label{local_env}
\end{figure*}

To quantify the environments of starburst galaxies in Illustris, we determine the number of galaxies with a minimum mass of 10$^5$M$_{\odot}$ (this is the minimum stellar mass of a resolved subhalo) surrounding our starburst and control samples within 500~kpc and 1~Mpc as shown in Fig.~\ref{local_env}. Using the same approach in section 3.1, we compose our control sample by selecting 500 random non-starburst galaxies. We find that the distribution in both samples peaks in the least dense environments, this confirms the findings of \cite{Wilkinson2017a} and \cite{Zabludoff1996} that suggest post-starbursts and therefore starbursts have a preference for low-density environments. We note that the pre-merger/harassment starburst sample has a relatively extended tail on the distributions suggesting that starbursts in denser environments are more likely to be driven by harassment events rather than mergers. The distribution of non-starburst galaxies is shifted towards denser environments meaning that starburst galaxies are in less dense environments than pre-merger/harassment starburst environments. We quantify this by performing a KS test and obtain a p-value much less than 1\% when comparing the pre-merger/harassment starburst sample to the post-merger sample at distances of 500~kpc and 1~Mpc.

When examining the halos our galaxies are in, we find the mean halo mass for the pre-merger/harassment starburst sample is 6.1$\pm$2.1$\times$10$^{12}$~M$_{\odot}$ compared with 3.4$\pm$2.0$\times$10$^{12}$~M$_{\odot}$ for the merger starburst sample. This result adds further evidence that starbursts without a previous merger are in denser environments to post-merger starbursts.

\subsection{Nuclear or Global Starbursts?}
In an extreme case, a starburst will use up all available cold gas throughout the galaxy, but as starbursts typically occur on a very short timescale, $\sim$50~Myr, these cases are very rare (\citealt{Mihos1994}). It is more likely that gas is funnelled into the galactic centre, triggering a nuclear starburst (\citealt{Barnes1991,Barnes1996}). It is unknown to what extent gas is consumed in this scenario because molecular hydrogen has been found in some post-starburst galaxies (\citealt{Zwaan2013}; \citealt{French2015}).

\begin{figure*}
	\centering
	\subfloat{
		\includegraphics[width=0.5\textwidth]{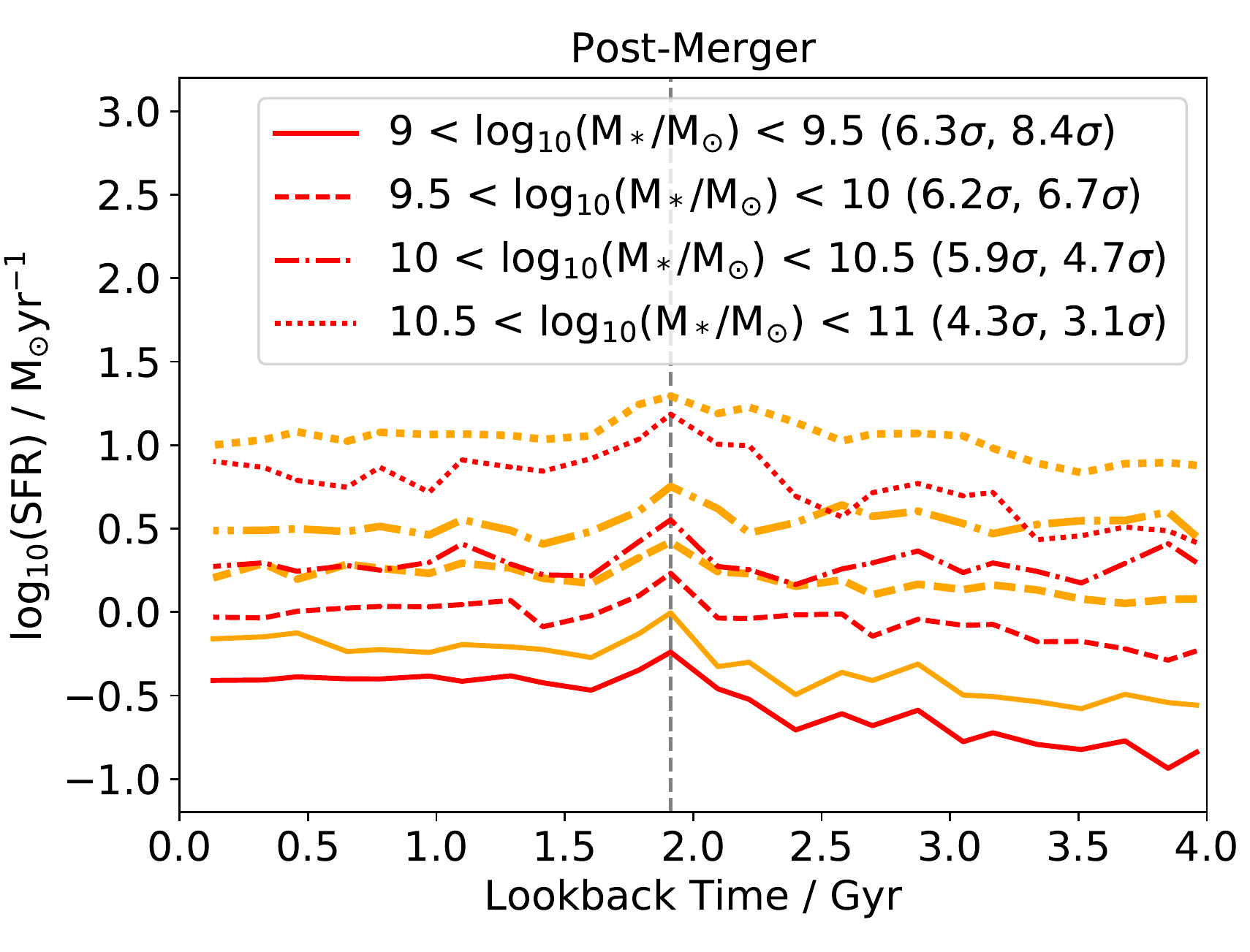}
	}
	\subfloat{
		\includegraphics[width=0.5\textwidth]{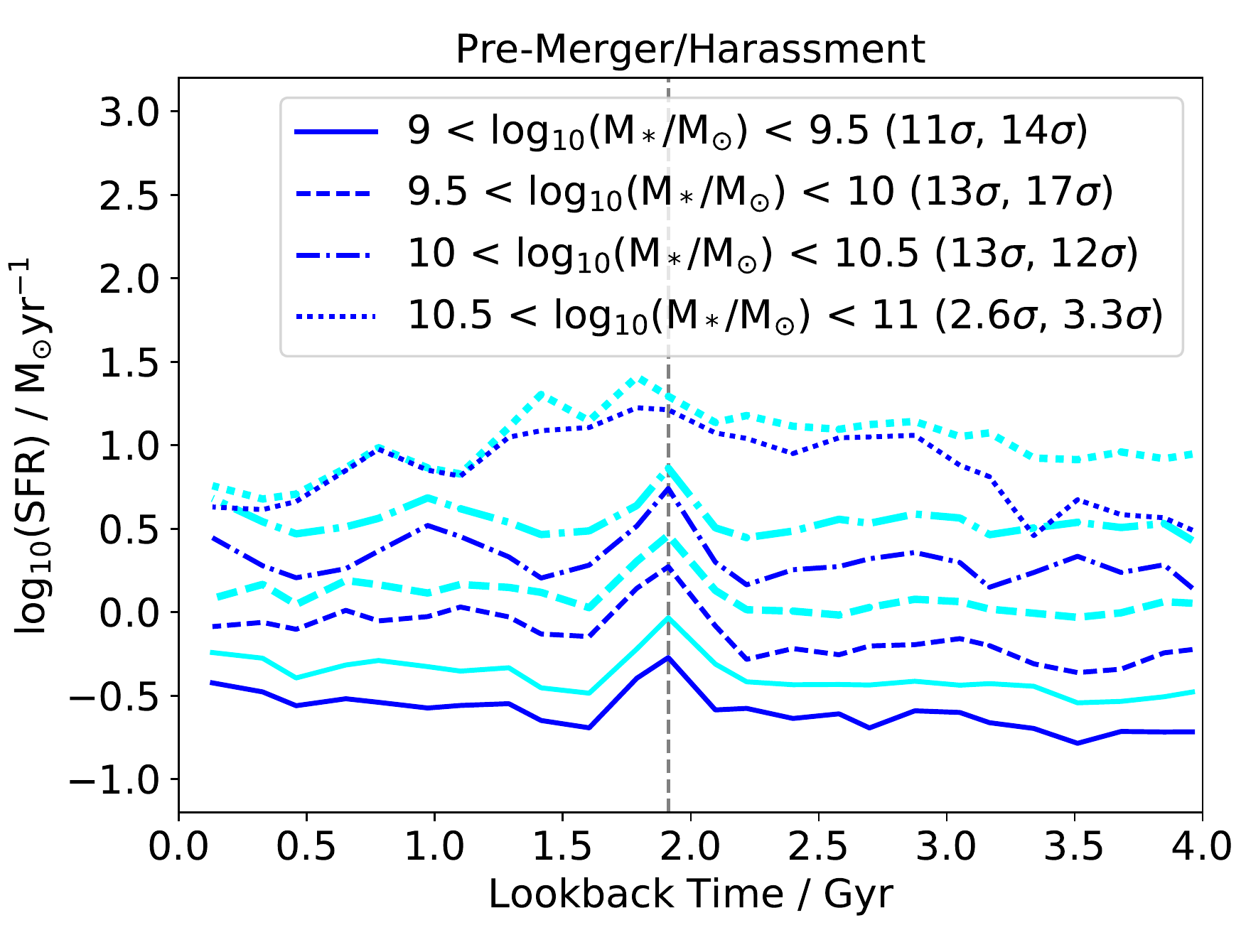}
	}
	\caption{Median stacked star formation rates (SFR) as a function of lookback time for post-merger starbursts (left) and pre-merger/harassment drivenrbursts (right). We include a grey dashed vertical line that denotes the temporal location of the starburst. We split our two samples by mass as given in Table~\ref{merger histories}. The lighter colours (orange and aqua) are correspondent to the total SFR, whereas the bolder colours (red and blue) are representative of SFR within the stellar half mass radius. In all cases the starburst can be witnessed as a peak in the middle of the plots. In the legends we include the significance of the peaks in terms of $\sigma$ (refer to main text for a description of how the significance is calculated) for the SFR in the stellar half mass radius and the total SFR respectively.}
	\label{sfr}
\end{figure*}

We start by examining the Star Formation Rates (SFR) of our samples (Fig.~\ref{sfr}) from $z$ = 0.36 to $z$ = 0.00 (this is 2~Gyr either side of the snapshot used for selection). We split each sample by the mass bins as defined in Table~\ref{merger histories}. We include the total SFR (orange and aqua) and the SFR within the stellar half mass radius (red and blue). In the legends we include the significance of the peaks in terms of $\sigma$ which is calculated by subtracting the median SFR before the peak from the height of the peak and dividing by the standard deviation of the pre-starburst SFRs. On average we find that SFR is higher in the pre-merger/harassment starburst sample than in the post-merger starburst sample. We don't see a significant difference between the total and stellar half mass radius.

\begin{figure*}
	\centering
	\subfloat{
		\includegraphics[width=0.5\textwidth]{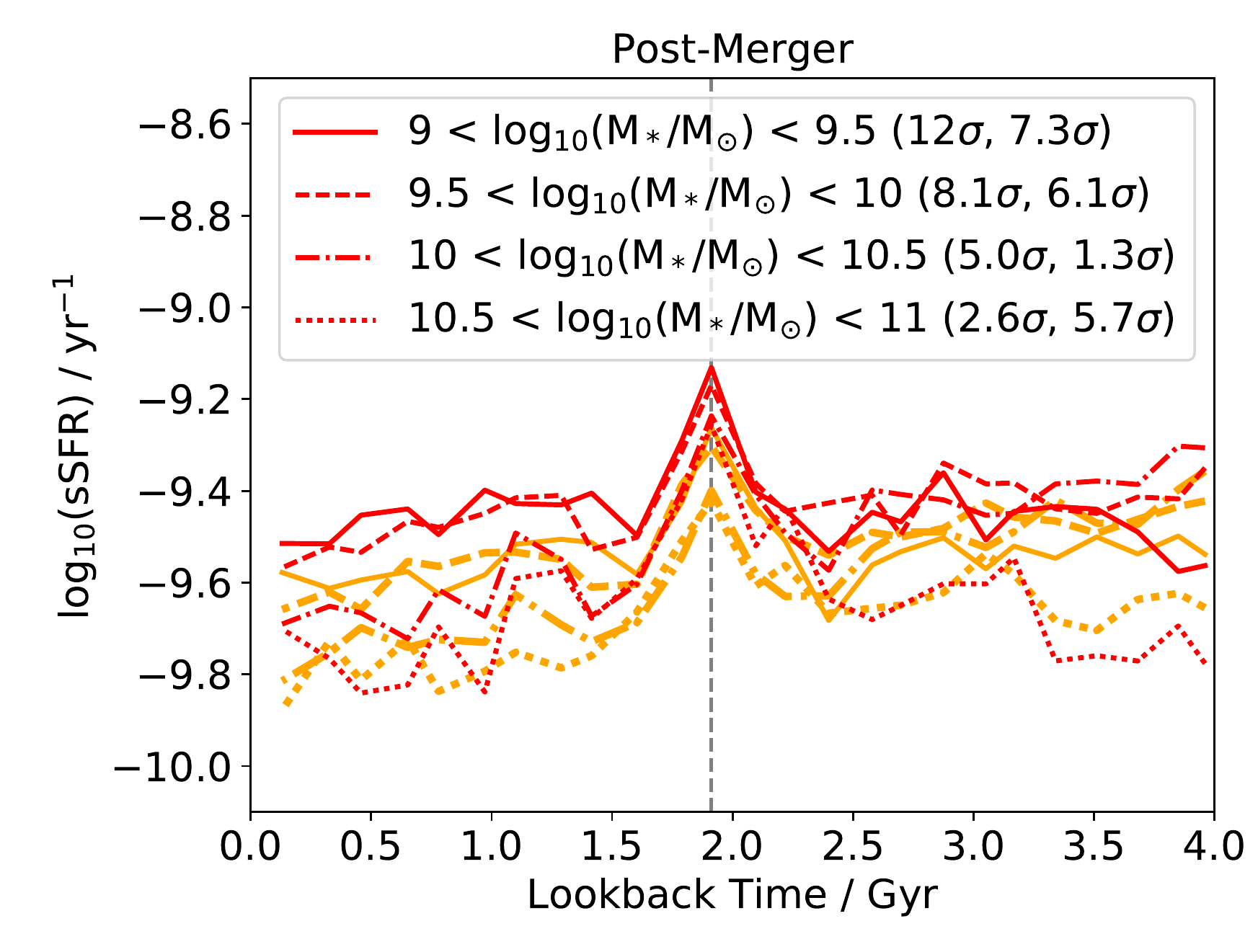}
	}
	\subfloat{
		\includegraphics[width=0.5\textwidth]{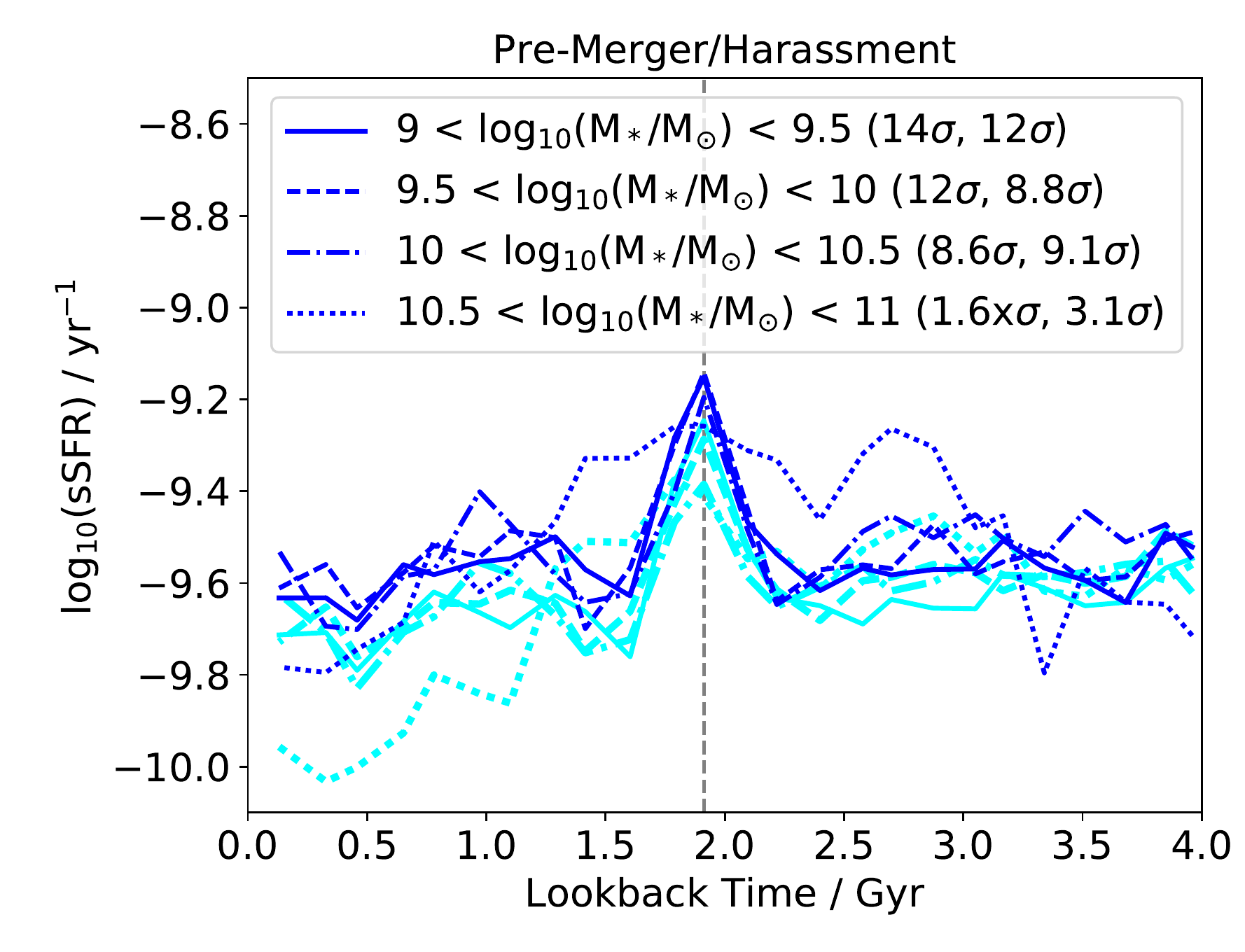}
	}
	\caption{Median stacked sSFR as a function of lookback time. The sSFR in both samples is higher in the stellar half mass radius when compared to the total radius. Again, we include the significance of all peaks within the legends for the stellar half mass radius and total radius respectively. A description of how the significant is calculated can be found in the main text.}
	\label{ssfr}
\end{figure*}

We also investigate the specific Star Formation Rates (sSFR) of our starburst galaxies. We plot sSFR against lookback time in Fig.~\ref{ssfr}, again split by mass bin and radii. In both samples we witness an offset between the total sSFR and sSFR within the stellar half mass radius, in which the sSFR in the stellar half mass radius is higher. When focusing on the enhancement at the time of the starburst, on average there is a higher enhancement in the stellar half mass radius. These results suggest that starbursts occur towards the galactic centre as noted by \cite{Barnes1991} and \cite{Barnes1996}.

We can also see that the enhancement in sSFR is 0.1~dex higher in the pre-merger/harassment starburst sample compared to the post-merger sample. However, this could be a consequence of the resolution of Illustris; this means that burstier star formation would be averaged out into the background star formation as the measure of time used to calculate star formation is greater at the time duration of the starburst. Therefore, this could be the result of post-merger starbursts being burstier than the pre-merger/harassment starburst which would explain a lower peak in star formation.

When mergers and interactions occur, tidal torques funnel gas into the galactic centre (\citealt{Barnes1991}, \citealt{Barnes1996}). This build up of gas then acts as a fuel for a rapid burst of star formation. In this section we explore the extent of this increase in gas and test whether there is an infall to the galactic centre.

\begin{figure*}
	\centering
	\subfloat{
		\includegraphics[width=0.5\textwidth]{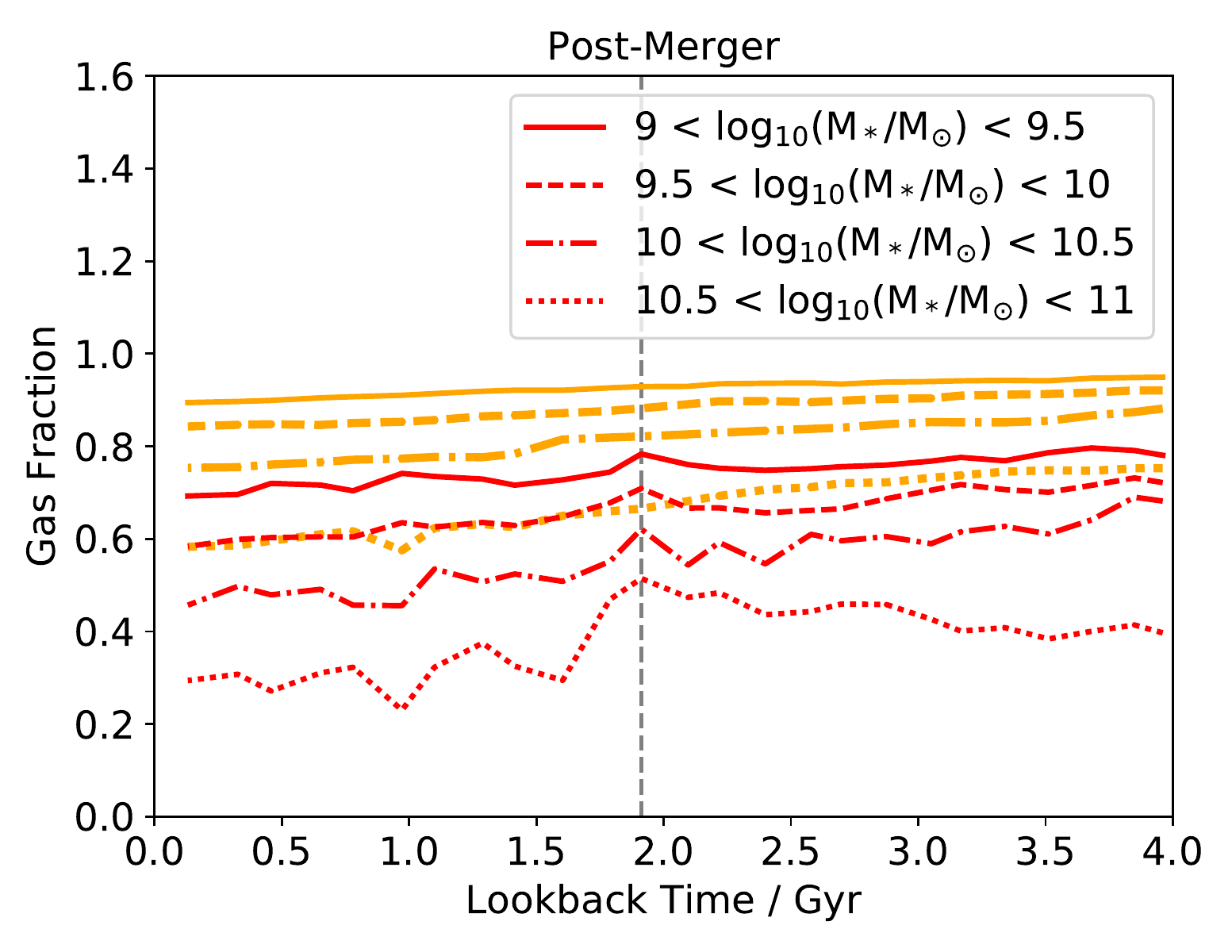}
	}
	\subfloat{
		\includegraphics[width=0.5\textwidth]{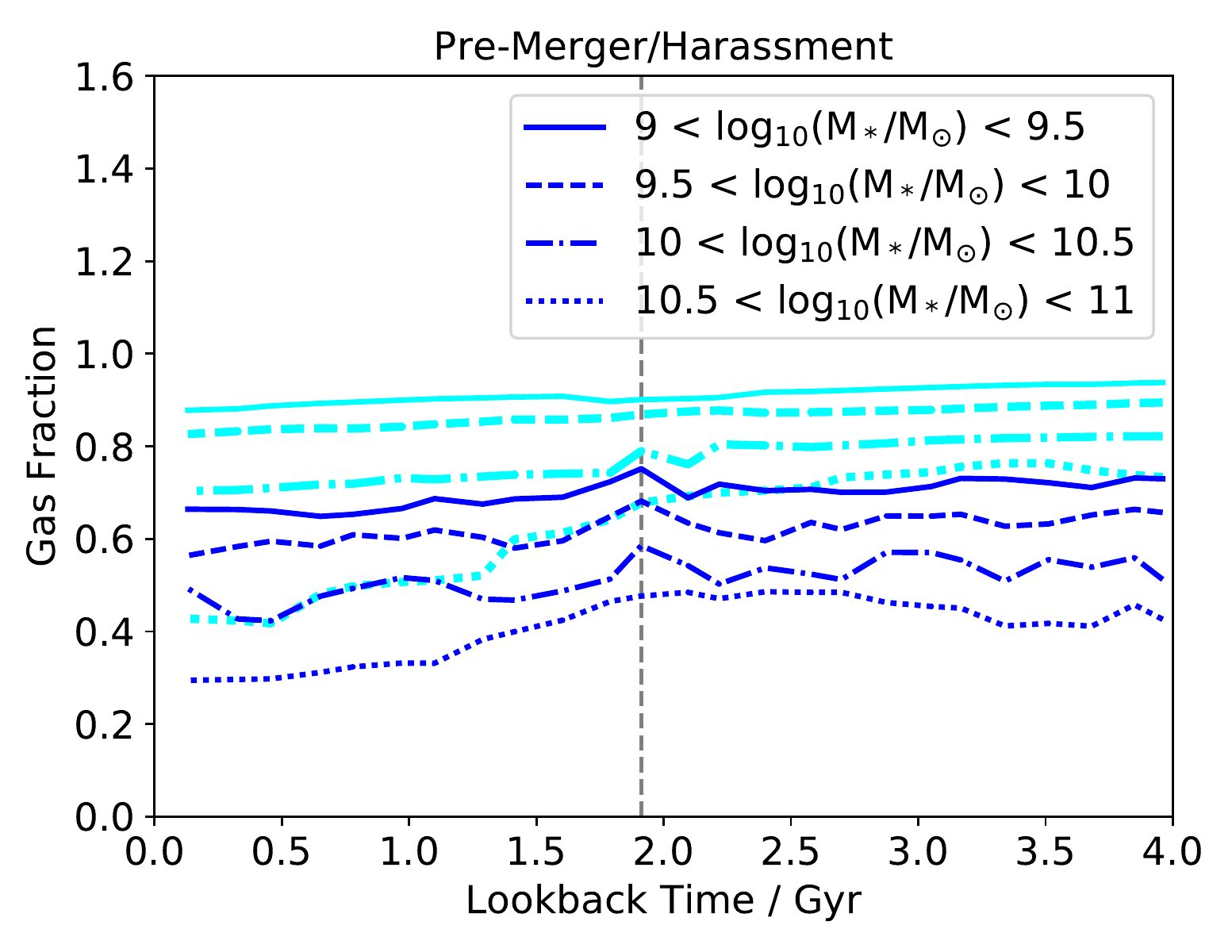}
	}
	\caption{Median stacked gas fractions against lookback time for post-merger starbursts (left) pre-merger/harassment triggered starbursts (right). Gas fraction is calculated as $M_{gas}$~/~($M_{gas}$+$M_{stars}$). There is no visible change in gas fraction within the total radius. However there is a slight but not significant change in gas fraction within the stellar half mass radius, this suggests star formation in Illustris is very efficient.}
	\label{gas_frac}
\end{figure*}

We plot the median stacked gas fractions within the stellar half mass radius and total gas fractions against lookback time in Fig.~\ref{gas_frac}. We calculate gas fraction to be $M_{gas}$~/~($M_{gas}$~+~$M_{stars}$), where $M_{gas}$ is the mass of the gas and $M_{stars}$ is the stellar mass. We see that there is a very mild downwards trend towards $z$=0 throughout each gas fraction calculated which is attributed to steady rates of star formation. Whilst we see no significant change in gas fraction within either sample at both the half mass and total radii, for both samples we see a slight fluctuation within the stellar half mass radius at the time of starburst. Due to the insignificance of this fluctuation, it could suggest that star formation in Illustris is very efficient. In the model described by \cite{Springel2003}, quiescent star formation rates increase with gas density, however if gas density surpasses a threshold value the gas consumption time scale becomes very rapid producing a burst of star formation. This increase in star formation efficiency in the galactic centre means that gas is quickly converted to stars. The insignificance of the peak in gas fraction could also be because the cold to hot gas ratio has increased during the starburst therefore increasing the star formation efficiency. We add the caveat that whilst it is possible to calculate the temperature of gas in Illustris, the resolution is not high enough to allow the probing of molecular gas clouds and therefore we are unable to locate the cold gas reservoirs in our galaxies.

By using a simulation that has shorter time intervals between snapshots, we may be able to witness a more significant rise and fall of the gas fraction within the stellar half mass radius. Because this fluctuation only appears in the half mass radii, this could indicate that the starburst event affects mainly the nuclear regions with a greater effect further afield only in extreme cases. A nuclear starburst would mean there is negligible change to gas fraction outside of the central region, which is what we see in Fig.~\ref{gas_frac}.

\subsection{Colour}
\begin{figure}
	\centering
	\includegraphics[width=0.5\textwidth]{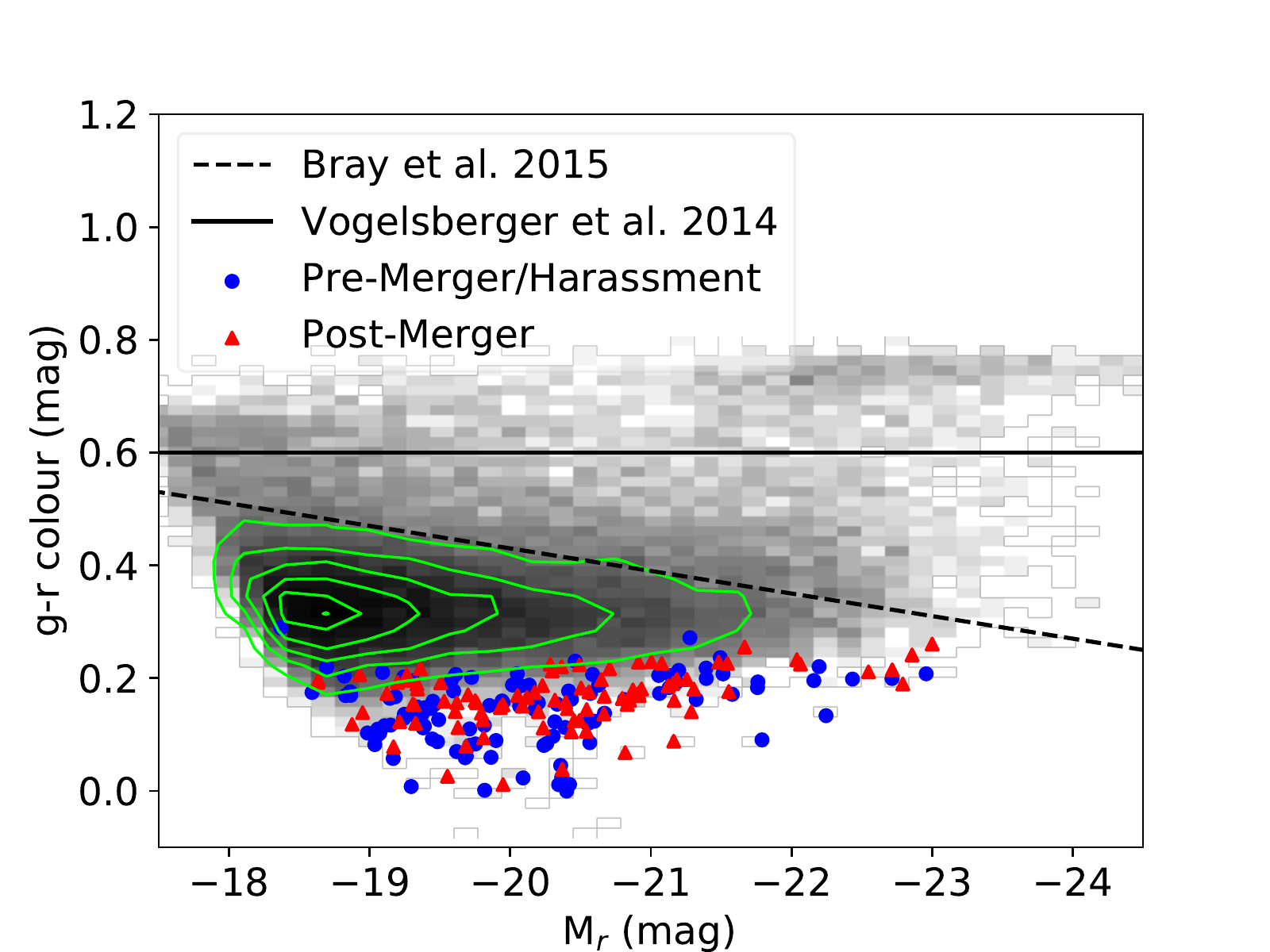}
	\caption{Colour-magnitude diagram at the time of starburst. We include the separation lines from \citealt{Bray2016a} (dashed line) and \citealt{Vogelsberger2014b} (solid line) that denote the locations of the blue cloud and red sequence. Illustris galaxies are denoted by the grey colour map and green contour lines. The darkest areas of the plot are the densest regions of the colour-magnitude diagram. It is clear from the distribution of Illustris galaxies, the red sequence is not well defined and hence the majority of galaxies reside in the blue cloud, this could be due to the absence of dust in the Illustris simulation. We find that both the post-merger (red triangles) and pre-merger/harassment (blue circles) starburst samples are located at the farthest regions of the blue cloud.}
	\label{colour_mag}
\end{figure}

\cite{Poggianti1999}, \cite{Poggianti2009} and \cite{Wilkinson2017a} find there is bimodality in colour when studying the properties of post-starburst galaxies. We explore this further by using the mock stellar photometry measurements Illustris provides. Fig.~\ref{colour_mag} shows a colour-magnitude diagram at the time of the starburst. We include the separation lines from \cite{Vogelsberger2014b} and \cite{Bray2016a} that denote the locations of the blue cloud and the red sequence. We add the caveat that the red sequence is not well defined in Illustris using photometry measurements, as there is no clear bimodality in colour. However, by plotting a colour-magnitude diagram, it allows us to determine whether one sample is bluer or redder than the other.

From Fig.~\ref{colour_mag}, we see that all starbursts are located well within the blue cloud region of the colour-magnitude diagram; situated under both the Vogelsberger and Bray lines (\citealt{Vogelsberger2014b} and \citealt{Bray2016a} respectively). The Illustris simulation does not include dust and therefore this is why there is no large range in the colour of starbursts. Instead, starbursts are `ultra-blue' in colour.

\begin{figure*}
	\centering
	\includegraphics[width=\textwidth]{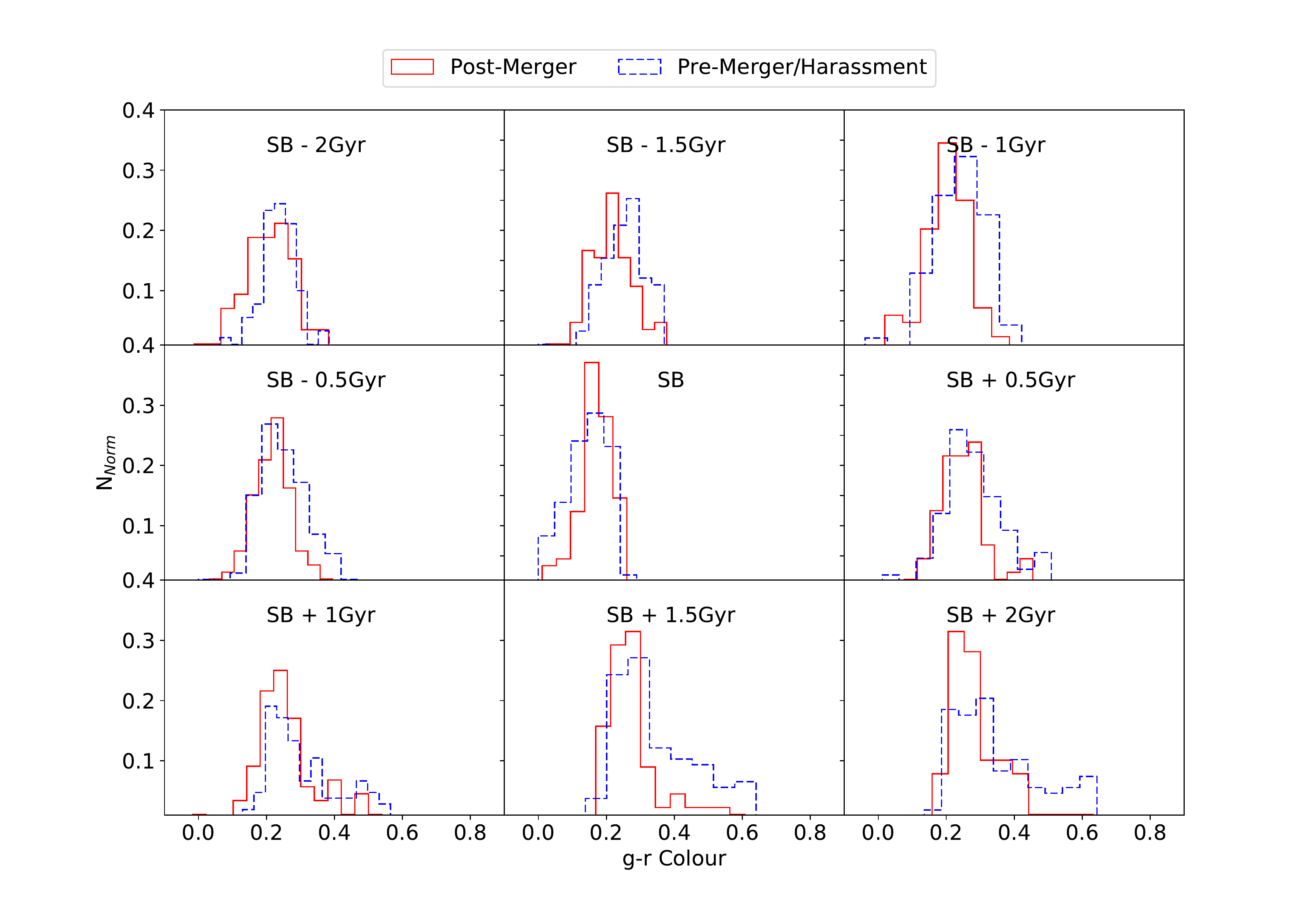}
	\caption{The g-r colour distributions 2~Gyr before and after the starburst event at intervals of 0.5~Gyr. We can see both populations are within the blue cloud and at the time of starburst (central plot) the distribution shift further into the blue cloud. After the starburst, the colour distributions shift to the right, slightly reddening.}
	\label{colour_hist}
\end{figure*}

We explore how colour changes over time in Fig.~\ref{colour_hist} by plotting the colour distribution at 0.5~Gyr intervals before, during and after the starburst event. We find that within the 4~Gyr range the vast majority of starbursts are below g-r = 0.6 and can be considered blue (\citealt{Vogelsberger2014b}). At the snapshot containing the starburst we can see that the colours of both samples become bluer and then shift to approximately their original positions after the starburst. However, we find that the pre-merger/harassment sample has an extended tail towards redder colours which becomes more pronounced as the galaxies progress through the post-starburst phase. We have seen previously in section 3.2. pre-merger/harassment starbursts have an extended tail into denser environments which suggests environment has an impact on the colour of starburst galaxies.

\subsection{Quenching and Feedback}
Active galactic nuclei (AGN) have been linked with post-starburst galaxies in many studies such as \cite{Trouille2011f}, \cite{Melnick2015} and \cite{Baron2017e} to name a few. It is heavily reported in the literature that mergers, particularly gas-rich major mergers, could be the main trigger behind AGN activity (\citealt{DiMatteo2005,Hopkins2006a}). This is the same trigger that is believed to ignite the starburst phase and hence post-starburst phase (\citealt{Zabludoff1996,Bekki2005a,Hopkins2006a}). In this section we briefly test to what extent AGN feedback plays in quenching star formation by examining black hole masses of the galaxies in this study.

\begin{figure*}
	\centering
		\subfloat{
		\includegraphics[width=0.5\textwidth]{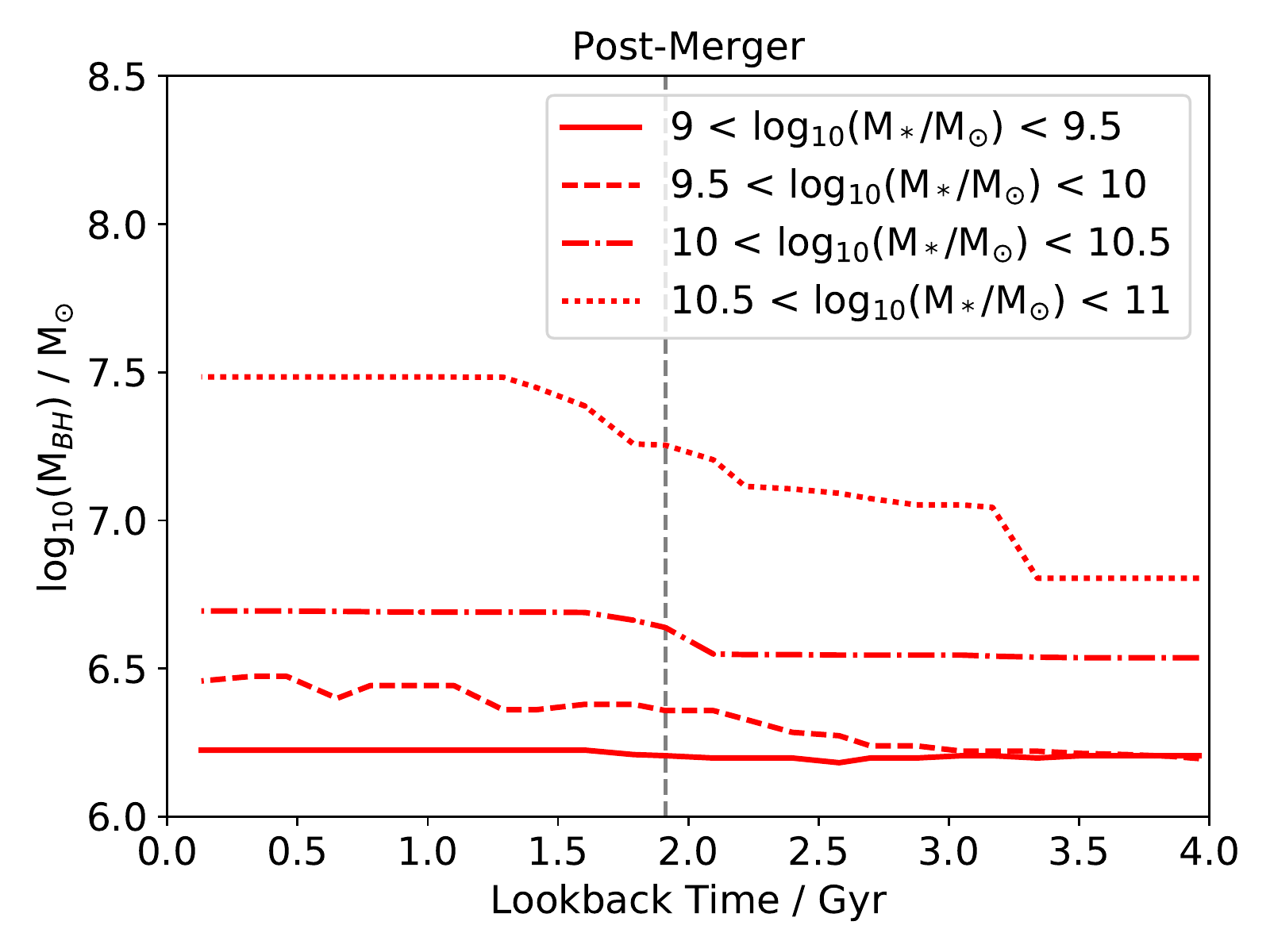}
	}
	\subfloat{
		\includegraphics[width=0.5\textwidth]{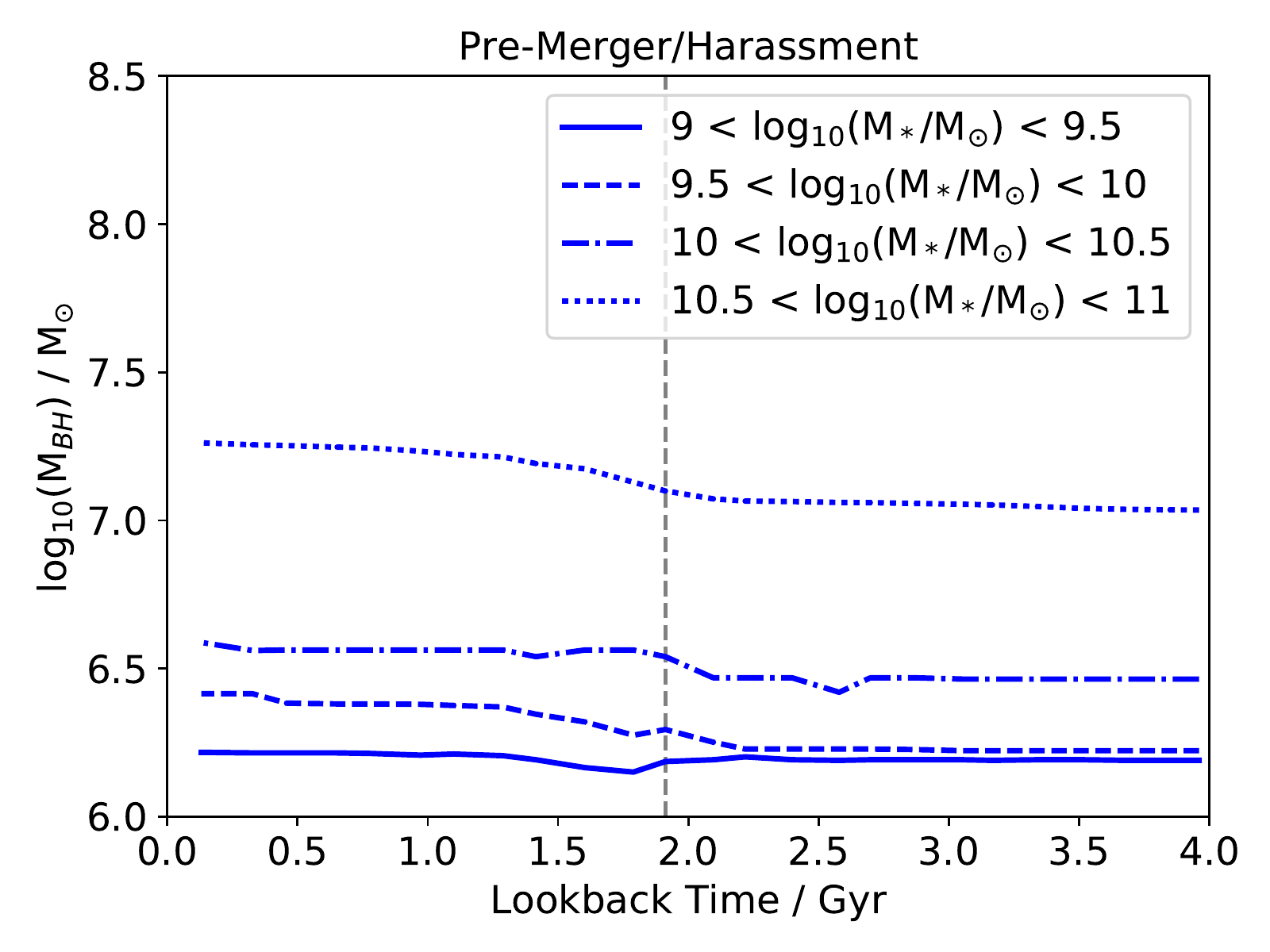}
	}
	\caption{Median black hole masses over lookback time for both samples. We have split each sample into the mass bins defined in Fig.~\ref{merger histories}. We see that for higher mass galaxies there is a slight gradual increase in black hole mass.}
	\label{bh_mass}
\end{figure*}

In Fig.~\ref{bh_mass}, we plot black hole mass against lookback time. Here, black hole mass is described at the sum of the masses of all blackholes in a subhalo. In both samples we see there is a gradual increase in black hole mass. For higher mass galaxies in the post-merger sample, there is on average a greater growth in blackhole mass which could indicate there is a larger level of AGN feedback, although there is no significant difference towards lower masses.

\section{Discussion}

Whilst previously thought that mergers are required to trigger a starburst, we find in this study, that harassment events can also trigger a starburst that is comparable in strength. We see that 55\% of the starbursts in this study have not had a previous merger in the past 2~Gyr. These starbursts appear to have been triggered by harassment events (that may or may not lead to a future merger), galactic instabilities (\citealt{Sparre2015}) or the accretion of gas from the surrounding intergalactic medium. Without the raw Illustris data we are unable to measure the tidal gravity between surrounding galaxies to test directly whether harassments are the main trigger for the pre-merger/harassment starbursts; this would be a suitable subject for future study. However, to get an indication of whether harassments could be a significant trigger, we test how close galaxies get to one another by studying their relative distances, as defined in Eq. \ref{eq_reldist}. We find that $\sim$40\% are within distances equivalent to the sum of their half mass radii, this means galaxies become very close, possibly to the extent in which their discs directly interact. At these distances it is likely that this could trigger a starburst.

In the literature, post-starburst galaxies have been shown to reside in low-density environments (\citealt{Zabludoff1996,Bekki2001c,Sanchez-Blazquez2009a}). We have seen that the starburst galaxies in this study are found to reside along the filament regions in low-density environments. When comparing the post-merger starbursts to the pre-merger/harassment starbursts we find that the pre-merger/harassment sample has a preference for denser environments. As denser environments have higher velocity dispersions, galaxies in close proximity are more likely to harass and fly-by than merge directly. We also compare the colours of both samples before, during and after the starburst and find the colour distribution of pre-merger/harassment starburst sample to be slightly redder than post-merger driven starbuples. We believe this is linked to environment and therefore starbursts that occur in denser environments are likely to be redder than low-density environments.

To examine the strength of the starbursts, we test SFR and sSFR and find that starbursts occur on timescale of $<$0.5~Gyr. With a higher resolution simulation and shorter time intervals between snapshot we would be able to make a more accurate measurement of starburst duration. Further, we see there are higher sSFR within the stellar half mass radius compared to the total sSFR. This suggests that starbursts occur more in the central regions as suggested by \cite{Barnes1991} and \cite{Barnes1996} rather than affecting the whole galaxy. This is also visible when we test gas fractions: while there is no significant enhancement in as fraction due to efficient star formation within the entire halo, we do witness a slight change within the stellar half mass radius which supports the nuclear starburst hypothesis.

We also find that the enhancement in sSFR at the time of the starburst is on average higher in the pre-merger/harassment starburst sample, which would suggest that starburst are stronger when not driven by a merger. However, due to the resolution of the Illustris simulation, it is more likely that starbursts in the merger driven sample are burstier than pre-merger/harassment starbursts and hence appear to be lower when averaged over time (\citealt{Sparre2016}).

We briefly investigate the extent AGN feedback could play in quenching star formation by measuring black hole masses over lookback time. We find there is on average more growth in black hole mass in the post-merger sample than there is in the pre-merger/harassment sample. This could indicate that there is more AGN feedback post-merger.

\section{Conclusions}
Starburst galaxies and the events that trigger them play an important role in transforming star forming spirals into quiescent ellipticals. The literature has many discrepant findings concerning the role of environment and triggering mechanisms. In this paper we have utilised the Illustris simulation to explore the possible triggering-mechanisms and making a comparison between post-merger and pre-merger/harassment triggered scenarios. We list here our principal findings:
\begin{enumerate}
	\item We find that 55\% of the starbursts identified in this study have not been triggered by a merger. The majority of this sample we believe to have been harassment driven due to their very close relative distances between surrounding galaxies, $<$~1.
	
	\item We find that in both of our samples, starburst galaxies are located within low-density regions in the filament regions of the cosmic web. The pre-merger/harassment driven starbursts have been found to have an extended tail in denser environments compared to post-merger starbursts.
	
	\item sSFR is on average larger within the stellar half mass radius which suggests a nuclear starburst rather than a galaxy wide starburst.
	
	\item Pre-merger/harassment starbursts have a slight extended tail towards redder colours in their colour distribution compared to post-merger starbursts. This is driven by environment and therefore denser environments produce redder post-starburst galaxies.
	
	\item These results suggest that mergers not only trigger bursty star formation but they could also trigger higher rates of feedback.
\end{enumerate}

These findings suggest that whilst there are two significant processes that can trigger a starburst of comparable strength, environment has an impact on which process a galaxies takes to enter the starburst phase. This also has an effect on the colour of the galaxy which in turn could contribute to the bimodality of colour of post-starburst galaxies we see in observational studies such as \cite{Poggianti1999}. Further work using the latest IllustrisTNG (\citealt{Nelson2017,Naiman2017,Springel2017,Pillepich2017,Marinacci2017}) will allow us to probe starburst galaxies in further detail with a higher temporal and spacial resolution.
 
\section*{Acknowledgements}
We thank the anonymous referee for their useful report that has helped improve the quality of this work. We would like to thank Sugata Kaviraj for his advice and useful discussions during this study. KAP and BKG acknowledge support of STFC through the University of Hull Consolidated Grant ST/R000840/1.
This research has made use of the Illustris database. The Illustris project acknowledges support from many sources: support by the DFG Research Centre SFB-881 ``The Milky Way System'' through project A1, and by the European Research Council under ERC-StG EXAGAL-308037, support from the HST grants program, number HST-AR-12856.01-A, support for program \#12856 by NASA through a grant from the Space Telescope Science Institute, which is operated by the Association of Universities for Research in Astronomy, Inc., under NASA contract NAS 5-26555, support from NASA grant NNX12AC67G and NSF grant AST-1312095, support from the Alexander von Humboldt Foundation, NSF grant AST-0907969, support from XSEDE grant AST-130032, which is supported by National Science Foundation grant number OCI-1053575. The Illustris simulation was run on the CURIE supercomputer at CEA/France as part of PRACE project RA0844, and the SuperMUC computer at the Leibniz Computing Centre, Germany, as part of project pr85je. Further simulations were run on the Harvard Odyssey and CfA/ITC clusters, the Ranger and Stampede supercomputers at the Texas Advanced Computing Center through XSEDE, and the Kraken supercomputer at Oak Ridge National Laboratory through XSEDE.




\bibliographystyle{mnras}
\bibliography{earef} 






\bsp	
\label{lastpage}
\end{document}